%% file: 20140426_Networks_Long.tex
\documentclass[a4paper, 11pt]{article}
\usepackage[T1]{fontenc}
\usepackage[utf8]{inputenc}
\usepackage{authblk}
\usepackage{graphicx}
\usepackage{amsfonts}
\usepackage{graphicx}
\usepackage{amsmath}
\usepackage{amstext}
\usepackage{comment}
\usepackage{subfig}
\usepackage{amssymb}
\usepackage{stmaryrd}
\usepackage[mathscr]{eucal}
\newcommand{\matr}[1]{\mathbf{#1}}
\newcommand{\tensor}[1]{\boldsymbol{\mathscr{#1}}}

\begin{document}
\hyphenation{Bi-coc-ca Po-li-gra-fi-ci}


\title{A Multiple Network Approach to Corporate Governance}

\author[1]{Fausto Bonacina}

\author[1]{Marco D'Errico\thanks{corresponding author: m.derrico2@campus.unimib.it}}

\author[2]{Enrico Moretto}

\author[1]{Silvana Stefani}

\author[3]{Anna Torriero}
\affil[1]{\small{Department of Statistics and Quantitative Methods, University of Milano -- Bicocca, Milan, Italy}}
\affil[2]{\small{Department of Economics, University of Insubria and CNR--IMATI, Milan, Italy}}
\affil[3]{\small{Department of Mathematics, Quantitative Finance and Econometrics, Catholic University, Milan, Italy}}


\date{ }

\maketitle

\abstract{\input{abstract.tex}}



\section{Introduction}
The structure and dynamics of complex systems has been one of the hottest research topics in the recent years (see for example \cite{Estrada2012}, \cite{BarBarVes2011}, \cite{Dorogo2010}, \cite{CohHav2010}, \cite{VegaRedondo2007}, \cite{AldLi2005}). A complex system can be described as a network whose nodes are the basic units of the system and the edges represent the link or relationship between the units, assuming that the edges embed all connections between nodes. Univariate link analysis concentrates on a single link type. For example, to analyze the structure of the Web, both HITS (Hyperlink--Induced Topic Search) (\cite{Klein1999}) and Page Rank (\cite{PageBrin1998}) decompose an adjacency matrix that represents the hyperlink structure. A review with some further results that deals with the comparison of such measures can be found in \cite{perrafortunato2008}.

However, the relationships between units can be many and of different type, generating a network for each attribute of the nodes. When many attributes are assigned to $n$ nodes, it becomes more and more important to analyze the resulting link structure, incorporating all links simultaneously. Tensor analysis is a technique to capture multilinear structures, when data have more than one mode. It is accomplished through matrix decomposition on a multi--way array. An increasing research stream nowadays is oriented towards the study and application of tensor analysis in system theory, signalling and circuit theory (\cite{KolBad2009}, \cite{KolBadKen2005}, \cite{DeLath}, \cite{Harsch1970}).

An adjacency tensor is formed by stacking the adjacency matrix for each attribute link to form a three way array. For multiple linkages, a three way array can be used, where the \textit{frontal slices} describe the adjacency matrix of each single attribute.

The concept of centrality can also be extended from the univariate to the multivariate case. In the univariate case, hubs and authorities values are reciprocally defined in a recursive scheme that is the basis of HITS algorithm (\cite{Klein1999}). In the multivariate case, the HITS model is generalized to the TOPHITS model (\cite{KolBadKen2005}). Moreover, hub and authorities find an interpretation related to the multiple linkages. Authorities and hubs are calculated accordingly through three--way Parallel Factor Analysis (PARAFAC), the higher order analogous of SVD (\cite{GolVanl1996}, \cite{KolBad2009}). In TOPHITS, the PARAFAC decomposition provides a three--way decomposition that yields hub, authority and attribute scores. The attributes that have the largest score are the most descriptive.

An alternative approach to analyze multi--way data is found in multiplex networks (see, for instance, \cite{Solaetal2013}, \cite{BatNicLat2013}). However, stacking the data in a matrix results in loss of information. Thus, the multiplex analysis is forcibly limited and requires more research to be effective (\cite{BatNicLat2013}). In \cite{bianconi2013}, the author proposes a statistical mechanics framework to describe multiple networks with  a significant overlap of the links in the different layers. In \cite{bianconi2013b}, the importance of using appropriate weighted measures of multiplex networks is stressed.

Tensor analysis is  employed in a variety of different fields. For example, bibliometric data can be analyzed through the body of work, distinguishing between papers written by different author  and predicting in which journal the paper is published (\cite{DunKolKeg}). Recent applications of higher order tensors have appeared in text analysis of web pages (\cite{KolBadKen2005}, \cite{KolBad2009}), spectroscopic fluorescence and wavelet-transformed multi-channel EEG (\cite{EvrBul}).

To our knowledge, tensor analysis has never been applied to economic or financial networks. We claim that the tensor approach can contribute substantially in the interpretation of Corporate Governance (CG thereafter), a central research topic in both theoretical and empirical financial literature. An exhaustive survey on relevant CG issues is presented in \cite{SV1997}. As it is well known, CG studies interrelations between ownership and control of a firm.
In the univariate case (i.e. considering only one network), many contributions can be found in the study of either Shareholding Network (SH) or Board of Directors (BD) network (\cite{BatCat2004}, \cite{DerSteTor},
\cite{Grassi2010}, and \cite{BR2006}). However, limited research has been done on the joint effect of the two networks considered together. For instance, in \cite{piccardi2010}, the authors find a remarkable overlap of the community structure in both networks. Stockholders invest their money in a company whose strategic decisions are taken by top management. Therefore, by analyzing simultaneously the two networks, tensor decomposition can contribute substantially to the interpretation of CG. This may give a unique perspective on the relevance of management rather than ownership in the company's policy.

In this paper, a multi--way analysis on SH and BD networks is performed on the CG multi--way array. In the CG multi--way array, hubs and authorities will also be found. The results will be contrasted with the univariate case.
Dominant subgroups will be found and interpreted. Italy is taken as a case study.

The paper is structured as follows. Corporate governance is introduced in Section 2, with a particular emphasis to Italy. Basic notions on graphs and tensor analysis are illustrated in Section 3. Description of data base and data
processing issues are in Section 4 while results are discussed in Section 5. A summary of results and conclusions are in Section 6.

\section{Corporate Governance}
Corporate Governance (CG) studies how to provide shareholders a certain degree of control on strategic actions undertaken by managers and how to measure their effectiveness and loyalty. It is crucial to detect which decisions might end up affecting, in a negative way, firms' market value and, consequently, shareholders wealth. Then a main goal of CG is to develop and propose practical rules that allow shareholders to enforce proper company management.

The analysis of CG in Italy is interesting for a number of reasons. First of all, recently, Italy issued rules on CG. In \cite{Melis1} and \cite{Melis2} the effect of the so--called 1998 ``Riforma Draghi'' whose aim was to give more power and financial protection to minority shareholders is analyzed. In particular, in most Italian listed companies the existence of a main shareholder that acts as a blockholder (i.e. a shareholder or a small group of shareholders that substantially influence company's decisions) can be found. Moreover, the Italian bank Mediobanca is linked, via many inter-corporate shareholdings, to most private groups.

Further, as of 2004, 60\% of listed companies were controlled by a shareholders' agreement (patto di sindacato) capable to put in the hands of a small groups of shareholders (either the state or some capitalistic families)
the full control of a company \cite{Melis2}. Just to mention a very recent event, in October 2013 the 'patto di sindacato' that controlled RCS Mediagroup (an important media company owner of 'Corriere della Sera', the most popular
Italian newspaper) was broken after having lasted since 1984. Involved in this agreement were (among others) the Agnelli family, Banca Intesa, Fondiaria-Sai and Assicurazioni Generali, that are very prominent players in the Italian
capitalist system. Furtermore, a separation between control and ownership is reachable through the so called \textit{pyramidal groups} if the actual number of held shares is relatively small \cite{Melis1}, \cite{KS1992}.

These ties might be at least partially affected by article 36 of the ``Decreto salva Italia'' that, since the beginning of 2012, prohibits interlocking (i.e. directors sitting in more than one board) between Italian banks, financial and insurance companies. The purpose of this rule is to avoid potentially anti--competitive behaviors and to reduce the usage of private financial information for speculative actions.

Within the European context, \cite{FL2002} analyzes companies' ownership in Western Europe, showing how many ``widely held'' or ``family controlled'' companies are present in financial markets. A comparison of the BD connections within Blue chips companies in Italy, France, Germany, United Kingdom, and United States is done in \cite{SDPG}. Through explanatory data analysis, it is established that the Italian case, as well as the French and German ones, is characterized by a small group of top managers sitting in several company boards, enforcing consequently strong ties between firms. Furthermore the Italian listed and unlisted companies are heavily characterized by a strong ownership concentration \cite{BC1999}.

In \cite{BB2006}, using descriptive analysis, a connection between corporate governance structure and firm performance in Italy is found. Even if under both a legal and an economical point of view between 1990 and 2005 there have been major changes into the Italian environment, only limited shifts toward more efficient control and ownerships structures can be detected. If minority shareholders do not have an adequate control on companies, blockholders can force managers to undertake business that are too risky ending up in a potential loss in company value. On the other hand, if most companies are connected in shareholding terms they can deploy collusive behavior. This, in the short run, can foster gains, but in the long run leads to a reduction in innovation and competitiveness.

In this paper, we analyze the complex structure of Corporate Governance of companies listed on the Italian Stock Exchange. To do this, we consider the tensor composed of the SH and the BD networks. This gives a unique approach to understand the relevance of ownership and management on the company's market performance. In this context, a question naturally arises. Are market price co--movements affected by a common CG structure? In order to provide a quantitative answer to this question, we consider a third network layer, the correlation network, obtained by constructing companies whose returns on price correlations is above a certain threshold ($0.65$).

\section{Methodology}
\subsection{Preliminaries}
We quickly recall some standard definitions and results about graph theory. For more details refer to \cite{Harary1969}. A graph $G=(V,E)$ consists of a set $V$ of $n$ vertices and a set $E$ of $m$ edges$.$ Let us denote with $\left\vert V\right\vert = n $ and $\left\vert E\right\vert = m $ the cardinality of the sets $V$ and $E$ respectively. A weight $w_{ij}$, $i, j = 1, \ldots, n$, is possibly associated to each edge $\left(  i,j\right)  $, in this case a weighted (or valued) graph is defined.

The degree $d_{i}$ of a vertex $i$ $(i=1,...,n)$ is the number of edges incident to it. A directed graph (digraph) is a graph in which all the edges are directed from one vertex to another. In a directed graph the in--degree of a vertex $i$ is the number of arcs directed from other vertices to $i$ and the out--degree of a vertex $i$ is the number of arcs directed from $i$ to other vertices.

A graph is connected if for each pair of nodes $i$ and $j,$ $\left(i, j = 1,2,...,n\right)  ,$ there is a path from $i$ to $j$. A digraph is strongly connected if for each pair of vertices there is a directed path connecting them.

The binary adjacency matrix of a graph $G=(V,E)$ is the nonnegative $n$-square matrix
$\mathbf{A}=\left[A_{ij}\right]  $, $\left(  i,j=1,2,...,n\right)  $
representing the adjacency relationships between vertices of $G$: the off--diagonal elements $a_{ij}$ of $\mathbf{A}$ are equal to $1$ if vertices $i$ and $j$ are adjacent, $0$ otherwise. Let $\left\{  \lambda_{1},\lambda
_{2},...,\lambda_{n}\right\}  $ be the set of the eigenvalues of $\mathbf{A}$ and $\rho=\max_{i}\left\vert \lambda_{i}\right\vert $ its spectral radius. If $G=(V,E)$ is undirected, it is known that $\mathbf{A}$ is symmetric and its eigenvalues are real. When $G=(V,E)$ is undirected and connected, $\mathbf{A}$ is irreducible, i.e. for every pair of indices $i$ and $j$, there exists a natural number $r$ such that $(\mathbf{A}^{r})_{ij}$ $\neq0$. If $G=(V,E)$ is
a digraph, its adjacency matrix $\mathbf{A}$ is in general asymmetric; moreover, $\mathbf{A}$ is irreducible if and only if $G$ is strongly connected. If $\mathbf{A}$ is a non--negative and irreducible matrix, it is well known that a positive eigenvector, called the Perron or principal eigenvector, corresponds to the spectral radius of $\mathbf{A}$.

In the case of a weighted graph without loops, i.e. without any edge connecting a vertex to itself, its adjacency matrix, denoted by $\mathbf{W}$ (the weighted adjacency matrix) has zero diagonal elements, and nonnegative real
off--diagonal entries.

Graphs with more than one type of link, also referred to as multigraphs or multiple networks, can be usefully described by tensors and their decompositions.

Two useful concepts for multiple networks analysis are the \textit{intersection} and \textit{union} networks. Given $K$ networks $G_{k}=(V,E_{k}),$ whose respective unweighted adjacency matrices are $\mathbf{A}_{k}=[A_{k}(i,j)]$, $k=1,\ldots, K$, the intersection network $G_{\cap}=(V,E_{\cap})$, is defined as follows: nodes $i$ and $j$ are linked in $G_{\cap}$ if and only  $\forall k$, $A_{k}(i,j) = 1$.
The adjacency matrix $A_{\cap}$ of the intersection network can be easily obtained through the Hadamard product $\mathbf{A}_{\cap} =  \mathbf{A}_{1}\circ \mathbf{A}_{2}\circ \ldots\circ \mathbf{A}_{K}$, that is the entry--wise product, between the $K$ adjacency matrices of the corresponding univariate networks. The union network $G_{\cup
}=(V,E_{\cup})$ is defined as follows: two nodes $i$ and $j$ are linked in $G_{\cup}$ if there exists a link between the two nodes at least in one of the $K$ networks. The adjacency matrix of the union network $\mathbf{A}_{\cup}$ is
obtained by summing up $\mathbf{A}_{1}, \mathbf{A}_{2}, \ldots, \mathbf{A}_{K}$ and binarizing the resulting
matrix. The union network allows to check the presence of any possible link between nodes.

In the following part of this section, some basic concepts about tensors are given  (for more details see \cite{KolBad2009}). \textit{Tensors} are multidimensional arrays. A multiple network, with $n$ nodes and $l$ link
types, can be represented by a three way (or third order) tensor $\boldsymbol{\mathscr{A}}$ of size $n\times n\times l$, whose the $(i,j,k)$ element, denoted by $A_{ijk}$, is different from zero if there is a link of type $l$ connecting node $i$ to node $j$. A \textit{slice} of a tensor $\boldsymbol{\mathscr{A}}$ is the matrix obtained by fixing the third index.
In particular, a \textit{frontal slice} of a tensor corresponds to the adjacency matrix of one type of network.

The \textit{Frobenius norm} of a tensor $\boldsymbol{\mathscr{A}}$ of size $n \times n \times l$ is defined as:
\begin{equation*}
\left\Vert \boldsymbol{\mathscr{A}}\right\Vert =\left(  \overset{n}%
{\underset{i=1}{\sum}}\overset{n}{\underset{j=1}{\text{ }\sum}}\overset
{l}{\underset{k=1}{\text{ }\sum}}a_{ijk\text{ }}^{2}\right)  ^{1/2}.
\end{equation*}

$\boldsymbol{\mathscr{A}}$ is a \textit{rank-one tensor} if it can be written as the \textit{outer product} of vectors. For instance, let $\boldsymbol{\mathscr{A}}$ be a rank--one tensor of size $n\times n\times l$, then $\boldsymbol{\mathscr{A}}=\mathbf{u}\circ\mathbf{v}\circ\mathbf{w}$, i.e. the $(i,j,k)$ element of $\boldsymbol{\mathscr{A}}$ is given by $A_{ijk} = u_{i} v_{j} w_{k},$ for $i,j=1,\ldots, n$, $k=1, \ldots, l$, where $\mathbf{u}\in \mathbb{R}^n, \mathbf{v} \in \mathbb{R}^n$ and $\mathbf{w} \in \mathbb{R}^n$

\subsection{HITS}\label{methodology}

This section illustrates the ways in which multiple networks can be analyzed in order to identify the key actors. We first recall the case of univariate networks, focusing in particular on centrality \cite{CrossPru2002} for
undirected networks and its univariate extension to the case of asymmetric, weighted and multiple networks.

A centrality measure enhancing the importance of a node as a function of centrality of the nodes to which it is connected, is the eigenvector centrality (see \cite{Bonac1987}, \cite{Bonac2007}): the centrality of an
actor is proportional to the sum of the centralities of the neighbors. Formally, the eigencentrality $x_{i}$ of the vertex $v_{i}$ is defined as $x_{i}=\alpha\sum_{j=1}^{n}A_{ij}x_{j};$ by setting $\alpha=\frac{1}{\rho}$, the value $x_{i}$ is represented by the $i$-th component of the Perron eigenvector $\mathbf{x}$ (see \cite{GraSteTor2007} for a review on eigencentrality and further results).

In general, the relations between actors are directed, so that the corresponding graph is a digraph and the adjacency matrix is asymmetric. Accordingly, the eigenvalue centrality has to be modified to take into account this asymmetry. Two attributes are given to each node: \textit{authority} and \textit{hubness} (\cite{Klein1999}). Authority measures prestige: actors who many other actors point to are called authorities. If a node has a high number of nodes pointing to it, it has a high authority value and this quantifies its role as a source of information. On the contrary, a hub is an actor referring to many authorities and its score measures acquaintance. Essentially, a good hub points to many good authorities and a good authority is pointed to by many good hubs.

More formally, let $G=(V,E)$ be the directed graph on $n$ nodes modeling the network and $\mathbf{A}(G)=[A_{ij}]$ the asymmetric adjacency matrix of $G$ of order $n$. Two scores (centralities) are associated to the $i-th$ node:
$a_{i}$ the authority score and $h_{i}$ the hub score.

These values can be reciprocally defined in a recursive scheme. This is the basis of the HITS algorithm (see \cite{Klein1999}). HITS leads to authorities and hubs computed iteratively as follows:
\begin{equation}
a_{j}^{(t+1)} = \sum_{i = 1}^n A_{ji}h_i^{(t+1)}
\label{auth score}
\end{equation}
\begin{equation}
h_{i}^{(t+1)}= \sum_{j = 1}^n A_{ij} a_{j}^{(t)}\label{hub score}
\end{equation}
or, equivalently,
\begin{equation*}
\mathbf{a=A}^{T}\mathbf{h}\text{ and }\mathbf{h=Aa},
\end{equation*}
where vectors $\mathbf{a}=[a_{1},\ldots,a_{n}]^{T}$ and $\mathbf{h}=[h_{1},...,h_{n}]^{T}$ give respectively the authority and the hub scores on all nodes. If the graph $G$ is undirected, i.e. the adjacency matrix is symmetric, then, for all nodes, authority and hub scores coincide.

Note that if the hub vector $\mathbf{h}$ is known with accuracy the authority scores can be computed from Eq. $\ref{auth score}$, and, analogously, for the hubs scores using Eq. $\ref{hub score}$. Since this is not the case, the idea of the HITS algorithm is to start with initial vectors $\mathbf{a}^{\left(  0\right)}$ and $\mathbf{h}^{\left(  0\right)}$ and to update hub scores and authority scores by repeating the process.

Note that authorities scores are computed by the current hub values, which in turn are computed from the previous authorities scores. It follows that hub and authority values are reciprocally defined in a recursive scheme. The convergence of HITS algorithm is guaranteed by the normalization of solutions after each iteration and by the condition $\lambda_{1}(\mathbf{A}^{T}\mathbf{A})>\lambda_{2}(\mathbf{A}^{T}\mathbf{A})$, being $\lambda_{1}$ and $\lambda_{2}$ the first two eigenvalues of $\mathbf{A}^{T}\mathbf{A}$. Denoting by $\mathbf{a}^{\left(k\right)}$ and $\mathbf{h}^{\left(k\right)}$ authorities and hub scores we get: $\mathbf{h}^{\left(1\right)} = \mathbf{A}\mathbf{a}^{\left( 0\right)}$, $\widehat{\mathbf{h}}^{\left(1\right)}=\mathbf{h}^{\left(1\right)} / \left\Vert \mathbf{h}^{\left(1\right)  }\right\Vert$, $\mathbf{a}^{\left(1\right)}=\mathbf{A}^T\widehat{\mathbf{h}}^{\left(1\right)}$ and, at the $t-th$ iteration, $\alpha\mathbf{h}^{\left(t\right)}
=\mathbf{A}^{T}\mathbf{Ah}^{\left(  t-1\right)  }$, $\alpha\mathbf{a}^{\left(
t\right)  }=\mathbf{AA}^{T}\mathbf{a}^{\left(  t-1\right)  }$, being $\alpha$ a normalization coefficient.

Performing power iteration method on $\mathbf{AA}^{T}$ and $\mathbf{A}^{T}\mathbf{A}$, $\mathbf{a}^{\left(t\right)}$ and $\mathbf{h}^{\left(  t\right)  }$ will converge respectively to the principal eigenvectors $\mathbf{a}^{\ast}$ and $\mathbf{h}^{\ast}$ of the symmetric semi-positive definite matrices $\mathbf{A}^{T}\mathbf{A}$ and $\mathbf{AA}^{T}$. If we consider the singular value decomposition (SVD) of $\mathbf{A}$, given by $\mathbf{A}=\mathbf{UDV}^{T}$, it is well known that the columns of $\mathbf{U}$ are the eigenvectors of $\mathbf{AA}^{T}$ and the columns of $\mathbf{V}$ are the eigenvectors of $\mathbf{A}^{T}\mathbf{A}$, called the left singular vectors and right singular vectors of $\mathbf{A}$, respectively. It follows that $\mathbf{a}^{\ast}$ and $\mathbf{h}^{\ast}$ correspond respectively to the principal right and left singular vectors of $\mathbf{A}$.
Finally, considering the first $K$ singular values of $\matr{A}$, i.e. $\sigma_{1}\geq\sigma_{2}\geq...\geq\sigma_{k}>0$, the matrix $\mathbf{A}$ can be approximated as follows \cite{GolVanl1996}:
\begin{equation*}
\mathbf{A}\approx\underset{i=1}{\overset{K}{\sum}}\sigma_{i}\mathbf{u}^{(i)}\circ\mathbf{v}^{(i)}
\end{equation*}
where $\mathbf{u}^{(i)}$ and $\mathbf{v}^{(i)}$ are the corresponding singular vectors. $\mathbf{x}%
\circ\mathbf{y}$ denotes the outer product of the two vectors $\mathbf{x}$ and $\mathbf{y}$, i.e. $\left(  \mathbf{x}\circ\mathbf{y}\right)  _{ij}=x_{i}%
y_{j}.$  The principal right and left singular vectors $\mathbf{a}^{\ast}$ and $\mathbf{h}^{\ast}$ corresponds to the best $rank-1$ approximation of $\mathbf{A}$ obtained for $k=1,$ so that $\mathbf{h}^{\ast}=\mathbf{u}^{(1)}$ and $\mathbf{a}^{\ast}=\mathbf{v}^{(1)}.$

\subsection{The TOPHITS algorithm}
When dealing with multiple networks, some techniques, known as \emph{tensor decompositions}, can be successfully
applied. Two major tensor decomposition methods are generally the most used: the PARAFAC (Parallel Factor Analysis, also called Canonical Decomposition, or CP) \cite{Harsch1970} and the Tucker decomposition (or TD, see \cite{Tucker1963}) from which many others are derived. The first one is a form of high order extension of SVD which allows to analyze three mode data yielding both hubs and authority scores and considering the context (the type of relation). Kolda and Bader \cite{KolBadKen2005} proposed the TOPHITS method which used the PARAFAC model to analyze a three--way representation of data from the Web.

The TOPHITS produces sets of triplets $\left\{  \mathbf{h}^{(i)},\mathbf{a}^{(i)},\mathbf{t}^{(i)}\right\}  $, where the $\mathbf{h}$ and $\mathbf{a}$ vectors represent hub and authority scores respectively and the $\mathbf{t}$ vector contains topic scores which identify the context of the links.

Analogously to HITS, these scores can be derived iteratively as follows:
\begin{equation}
a_{j}^{(t+1)}=\sum_{i = 1}^n \sum_{k = 1}^l A_{jik} h_i^{(t+1)} t_k^{(t)}, \qquad  j=1,\ldots,n;
\end{equation}
\begin{equation}
h_{i}^{(t+1)}=\sum_{j = 1}^n \sum_{k=1}^l A_{ijk} a_j^{(t)} t_k^{(t)}, \qquad i=1,\ldots,n;
\end{equation}
\begin{equation}
t_{k}^{(t+1)}= \sum_{i = 1}^n \sum_{j = 1}^n A_{ijk} a_j^{(t+1)} h_i^{(t+1)}, \qquad k=1,\ldots,l.
\end{equation}

Following the notation in \cite{KolBadKen2005}, let $\boldsymbol{\mathscr{A}}$ denote the $n\times n\times l$ adjacency tensor of the multiple networks. Let $A_{ijk}=1$ if there is a relationship of type $k$ from actor $i$ to actor $j$ , $A_{ijk}=0$ otherwise. If a weight is associated to each edge $(i,j)$ in the $k$--th network, then we set $A_{ijk}=w_{ij}^{(k)}$.

The above equations can be written as:
\begin{align}
&  \mathbf{h}^{\left(  t+1\right)  }=\boldsymbol{\mathscr{A}}\text{ }%
\overline{\times}_{2}\mathbf{a}^{\left(  t\right)  }\overline{\times}\text{
}_{3}\mathbf{t}^{\left(  t\right)  }\label{pf1}\\
\bigskip & \nonumber\\
&  \mathbf{a}^{\left(  t+1\right)  }=\boldsymbol{\mathscr{A}}\overline{\times
}_{1}\mathbf{h}^{\left(  t+1\right)  }\overline{\times}_{3}\mathbf{t}^{\left(
t\right)  }\label{pf2}\\
& \nonumber\\
&  \mathbf{t}^{\left(  t+1\right)  }=\boldsymbol{\mathscr{A}}\overline{\times
}_{1}\mathbf{h}^{\left(  t+1\right)  }\overline{\times}_{2}\mathbf{a}^{\left(
t+1\right)  } \label{pf3}%
\end{align}
where $\boldsymbol{\mathscr{A}}$ $\overline{\times}_{r} \mathbf{a}\overline{\times}_{q}\mathbf{t}$ indicates that the tensor $\boldsymbol{\mathscr{A}}$ multiplies the vectors $\mathbf{a}$ and $\mathbf{t}$ in dimensions $r$ and $q$ respectively. For example, if $r=3$ and $q=2$, the $i-th$ component of $\mathbf{s}=\boldsymbol{\mathscr{A}}$
$\overline{\times}_{3}\mathbf{a}\overline{\times}_{2}\mathbf{t}$ is
\begin{equation*}
s_{i}=\underset{k=1}{\overset{m}{\sum}}\underset{j=1}{\overset{n}{\sum}}A_{ijk}a_{k}t_{j}\text{ \ \ \ for \ \ } i = 1,...,n.
\end{equation*}

Analogously to the HITS algorithm based on the SVD matrix decomposition, the TOPHITS algorithm based on the PARAFAC tensor decomposition reveals latent groupings of nodes that can be associated to each factor $r$, but TOPHITS also
includes latent information related to the network layer (\cite{KolBadKen2005} for a thorough description).

Using the PARAFAC decomposition \cite{Harsch1970}, the tensor $\tensor{A}$ can be approximated as follows:
\begin{equation}
\label{PARAFAC} \tensor{A}\approx\underset{r=1}{\overset{R}{\sum}}\lambda_{r}%
\mathbf{u}^{(r)}\circ\mathbf{v}^{(r)}\circ\mathbf{w}^{(r)}%
\end{equation}
where $\mathbf{x}\circ\mathbf{y}$ $\circ$ $\mathbf{z}$ denotes the three way
outer product of the three vectors $\mathbf{x}$, $\mathbf{y}$ and $\mathbf{z}$
i.e. $\left(  \mathbf{x}\circ\mathbf{y}\circ\mathbf{z}\right)  _{ijk}%
=x_{i}y_{j}z_{k}$, and $\lambda_r \in \mathbb{R}^R$ is a scaling factor (see \cite{KolBad2009} for a detailed discussion).

As before, under proper hypotheses, Equations \ref{pf1}, \ref{pf2} and \ref{pf3} converge to the best $rank-1$ approximation of $\mathbf{A}$ obtained for $k=1,$ so that $\mathbf{h}^{\ast}=\mathbf{u}^{(1)}$ and $\mathbf{a}^{\ast
}=\mathbf{v}^{(1)}$ and $\mathbf{t}^{\ast}=\mathbf{w}^{(1)}$.

The PARAFAC can be seen as a special case of TD, which is a higher--order extension of Factor Analysis and Principal Component Analysis (PCA). The TD features a \textit{core tensor}, whose entries indicate the level of interaction between the various components. More precisely, in the three way case, using the TD, a tensor $\boldsymbol{\mathscr{A}}$ of size $I\times J\times K$ can be approximated as follows (for more details see \cite{KolBad2009}):
\begin{align}
\boldsymbol{\mathscr{A}}  &  \approx\underset{p=1}{\overset{P}{\sum}}%
\underset{q=1}{\overset{Q}{%
{\displaystyle\sum}
}}\underset{r=1}{\overset{R}{%
{\displaystyle\sum}
}}\underset{}{\overset{}{g_{pqr}}}\mathbf{u}^{(p)}\circ\mathbf{v}^{(q)}%
\circ\mathbf{w}^{(r)} =\nonumber\\
&  = \boldsymbol{\mathscr{G}}\times_{1}\mathbf{U}\times_{2}\mathbf{V}%
\times_{3}\mathbf{W}\nonumber
\end{align}
or, elementwise:
\begin{align}
&  A_{ijk}\approx\underset{p=1}{\overset{P}{\sum}}\underset{q=1}{\overset{Q}{%
{\displaystyle\sum}
}}\overset{R}{\underset{r=1}{%
{\displaystyle\sum}
}}G_{pqr}u_{ip\text{ }}v_{jq}w_{kr}\label{TUCKER}\\
&  \text{for \ \ }i=1,\ldots,I,\text{ \ \ }j=1,\ldots,,J,\text{ \ \ }%
k=1,\ldots,K\nonumber
\end{align}
where $\mathbf{U}\in\mathbb{R}^{I\times P}$, $\mathbf{V}\in\mathbb{R}^{J \times Q}$ and $\mathbf{W}\in\mathbb{R}^{K\times R}$ are the component matrices and $\boldsymbol{\mathscr{G}}\in\mathbb{R}^{P\times Q\times R}$ is called the \textit{core tensor}. $\boldsymbol{\mathscr{G}}\times_{n}\mathbf{U}$ denotes the $n$ mode product of the tensor $\boldsymbol{\mathscr{G}}$ with the matrix $\mathbf{U.}$ Note that the Tucker Decomposition is not unique and the component matrices are usually orthogonal.

\section{The tensor of Corporate Governance in the Italian financial market}
Companies in a financial market can be connected in many ways, e.g. sharing a common industrial sector or having a similar price dynamics. Doubtless, the most relevant connection is given by the corporate control. Holding companies
possess an influential control in term of corporate governance on the controlled companies. Corporate governance is the system by which companies are directed and controlled and involves a set of relationships between a company's management, its board, its shareholders and other stakeholders \cite{Cadb1992}. We examine two levels of such a system: the shareholding structure and the board structure in the year 2013 in Italy. Both levels imply a pairwise relationship between companies, thus leading to two networks: the shareholding network (SH) and the interlocking board of
directors network (BD) (\cite{Grassi2013}, \cite{RotDar2013}, \cite{DerSteTor}). We consider the dataset of $273$ quoted companies on the Italian Stock Market (sources: Borsa Italiana and Bloomberg).

Networks along the tensor are defined as $A(: , :, k)$, where the index $k$ stands for the $k$--th network in the system (or, alternatively, the $k$--th layer in the multiple network). Figure \ref{fig:fin_mult} shows the multiple network structure represented as a tensor:
\begin{figure}[tbh]
\centering
\vspace{0.5cm}
\includegraphics[scale=0.36]{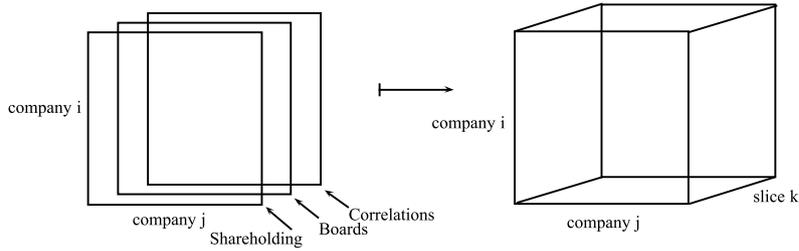}\caption{The
Corporate Governance multiple network represented as a tensor}%
\label{fig:fin_mult}%
\end{figure}
\subsubsection*{Shareholding network (SH)}
The first layer/network is the Shareholding Network (SH), defined as follows:
\begin{equation*}
A(i, j, 1) = \text{percentage of (quoted) company $j$ held by (quoted) company $i$}.
\end{equation*}

All relative amounts below the threshold of 2\% are discarded. 2\% is the limit imposed by the Italian \emph{Commissione Nazionale per le Societ\`{a} e la Borsa} (CONSOB) to consider, for a listed company, a relevant shareholder. Any
subject holding more than 2\% of a listed company must inform CONSOB. The SH network is weighted and directed, with rational weights between $0$ and $1$ \cite{DerSteTor}.

The SH network considers quoted holdings only. In the Italian market, a significant part of the control structure belongs to non listed companies. In fact, several non--quoted companies hold quotas of quoted companies. Hence
pairs of quoted companies might be linked via an external relationship, where the common holding company does not appear. On the other hand, at the board level it is more frequent that commonly held companies (via an external one)
share a significant number of directors. Thus, we find extremely useful to analyze simultaneously both networks, in order to find common patterns in the CG structure. Another feature of the Italian market is that a strong
government/public control exists in several quoted companies. The Ministry of Finance and the former Postal Service hold important quota of very large high--cap companies (e.g. Cassa Depositi e Prestiti vs ENI and Terna, and
the Ministry of Finance vs ENI and ENEL) as well as several municipalized companies (e.g. Comune di Milano, Roma Capitale, Comune Reggio Emilia and many others). In this case, the holders are not quoted and a significant quota of
information is missing.

\subsubsection*{Board of directors network (BD)}
The second network/layer is represented by the board of directors network, defined as
\begin{equation*}
\text{$A(i, j, 2)$ = \# of directors sitting in both $i$ and $j$}%
\end{equation*}

The univariate BD network has been studied extensively \cite{BatCat2004} and, more recently, \cite{Grassi2010}
and \cite{Grassi2013}. The adjacency matrix contains all listed companies in the Italian Stock Exchange for the year 2013. Isolated vertices have been excluded. The network is undirected  and weighted with integer weights. Moreover, it is not connected, characterized by a giant component and the presence of a few hubs.

\subsubsection*{Correlation network}
In addition to this, it is interesting to wonder whether similar governance structure lead to similar co--movements in market prices. In order to quantitatively assess this, we created an additional network layer, representing the correlations of price returns. In particular, we built a binary network that takes values $1$ when the value of the return correlation coefficient is bigger than $0.65$. In this way, we want to establish a linkage between a pair of companies only when a significant positive correlation in the price structure occurs.

The third network/layer is hence the binary correlation network defined as:
\begin{equation*}
X(i, j, 3) = \left\{
\begin{array}{c l}
    1 & \text{ if the value of the correlation coefficient} \\
    &  \text{of price returns
is bigger than }0.65 \\
& \\
    0 &\text{ otherwise.}
\end{array}\right.
\end{equation*}

This network has been introduced to assess and quantify the effect on the Corporate Governance structure of the movements in market prices.
\begin{figure}
\begin{minipage}{.5\linewidth}
\centering
\subfloat[Layer 1 -- SH network]{\label{main_multiple:a}\includegraphics[scale=.26]{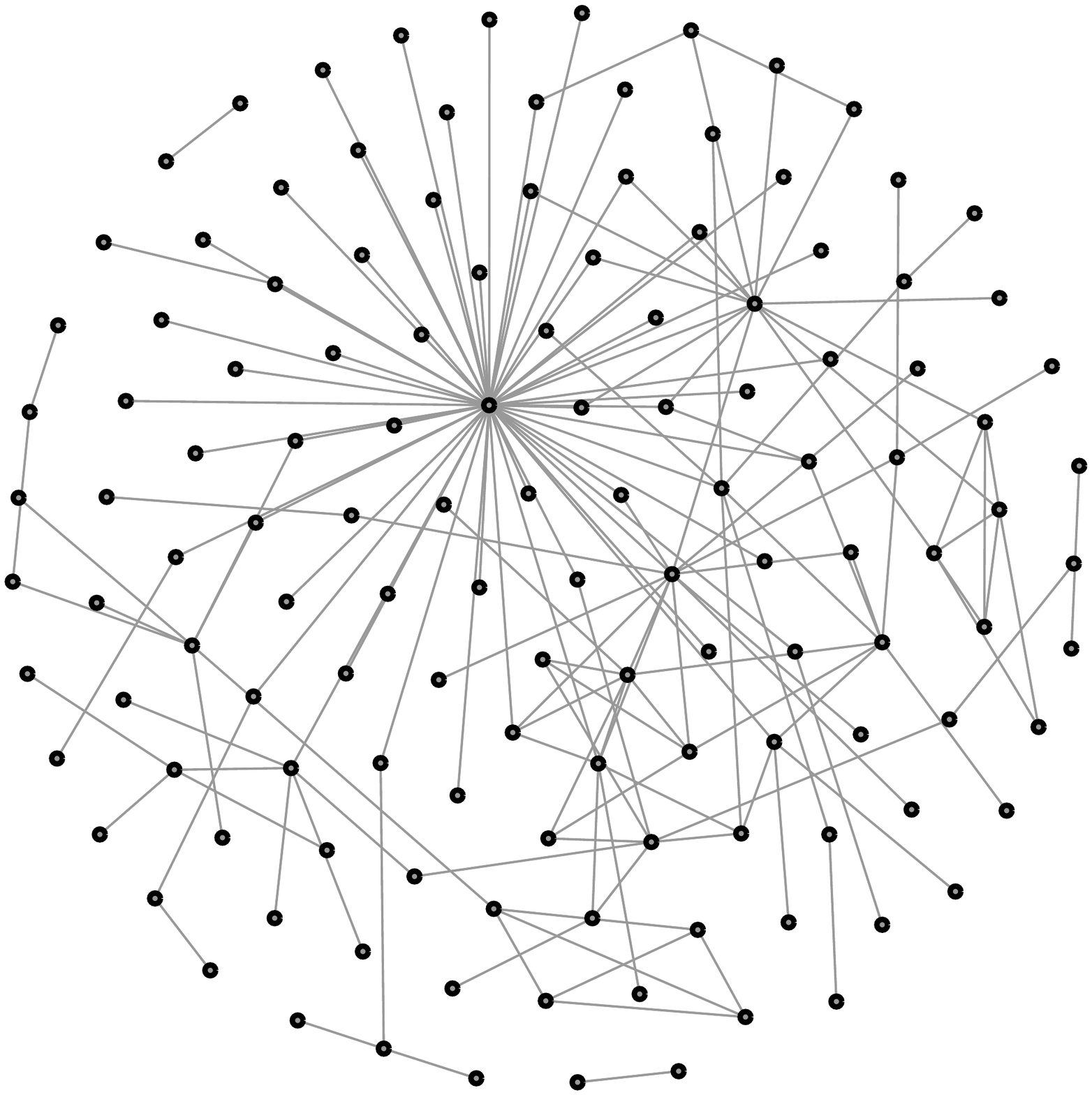}}
\end{minipage}%
\begin{minipage}{.5\linewidth}
\centering
\subfloat[Layer 2 -- Board network]{\label{main_multiple:b}\includegraphics[scale=.26]{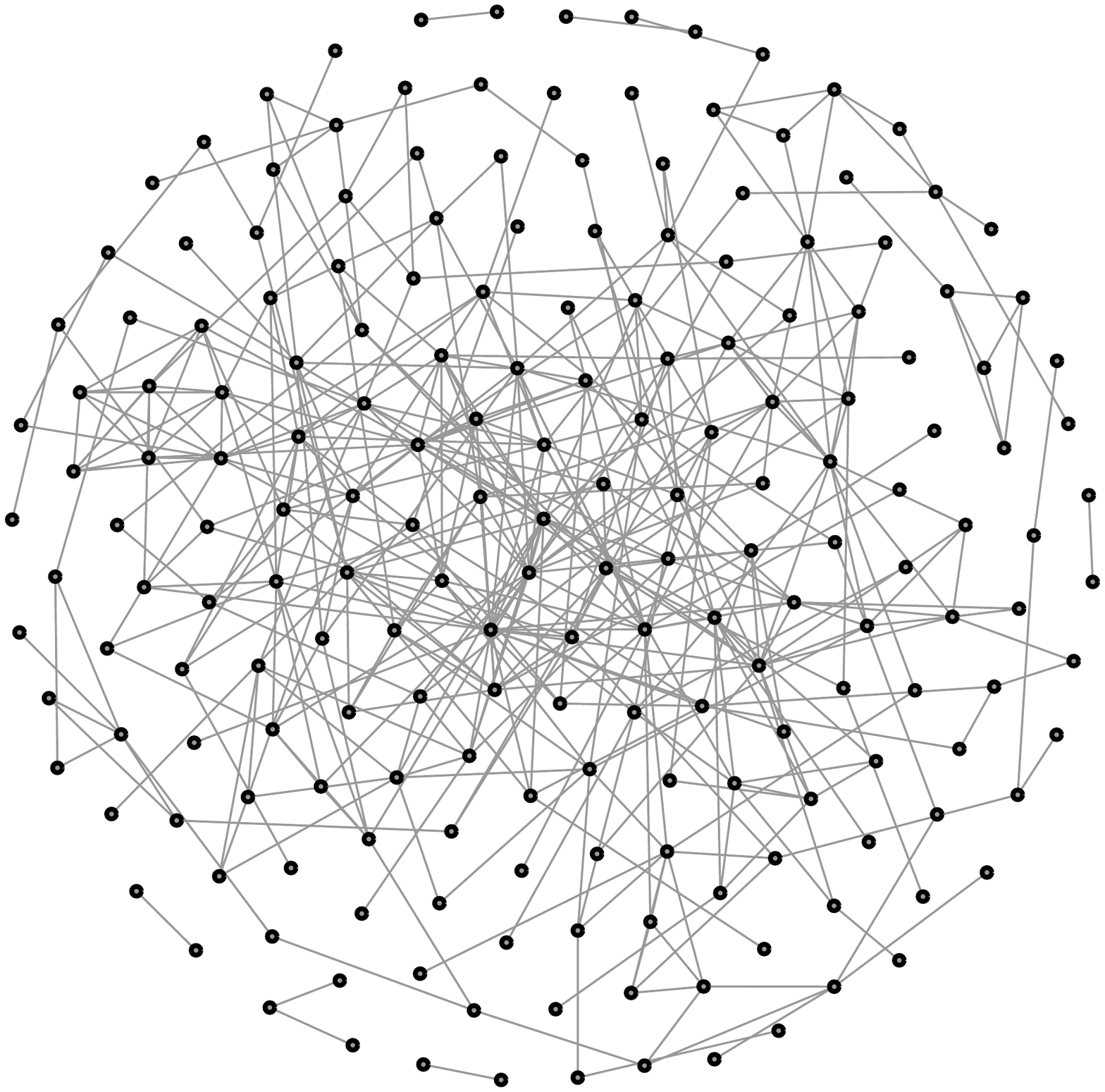}}
\end{minipage}\par\medskip
\caption{The SH network and the BD network (with no isolated nodes) representing the layers of the Italian CG multiple network}
\label{fig:main_multiple}
\end{figure}

\subsection{Empirical data}
We applied our model to Italian data. In particular, we consider 273 companies listed in the FTSE--Italia All Share index. Data are collected from Bloomberg database. To construct the two networks, for all the above mentioned companies, we considered the board of directors composition as of July 2013 and the structure of shareholdings as of 16 July 2013. Finally, in order to obtain the correlation network, we considered the time series of daily closing prices starting from 2 May 2012 and ending at 30 April 2013 for each company.

All computations have been done by using the Matlab Tensor Toolbox Version 2.5 \cite{KolBad2012} and the Matlab N--Way toolbox \cite{AndBro2000}. Both toolboxes provide very similar results, thus showing a certain degree of numerical robustness of the method proposed.

\subsection{Data pre--processing}
Given the particular nature of the methodology hereby proposed, data had to be pre--processed (see, for example for a discussion on this problem \cite{KolBadKen2005}). In particular, we had to develop a heuristics to deal with three main problems:
\begin{itemize}
\item while applying the PARAFAC algorithm to compute the TOPHITS scores for hubs/autho--rities/layer score, an appropriate number of factors has to be found that would guarantee a good fit while avoiding overfitting at the same time.
\item given the nature of the PARAFAC algorithm, we wanted to deal with a multiple network whose union was strongly connected and intersection was non--empty. In order to achieve this, we reduced the nodes to $182$. This allowed to find a sub-network where the union network is strongly connected.
\item creating a homogeneous weighting structure for both SH and BD. In fact, since these two networks are different in terms of both topological structure and network density, we needed to look for a way to make them comparable in topological terms.
\end{itemize}

In order to solve the first and third problem, after several attempts, we found a heuristic rule that seemed to work well in this context. At first, we discarded both the self--owned shares in the SH network, thus setting to
zero the main diagonal of the first slice (we posed the main diagonal of the second slice to zero as well). Secondly, in order to make it possible to obtain a reasonable comparison of the topological structure of the two layers,
we normalized each layer by its Frobenius norm. 


Moreover, the lack of a well established methodology for the determination of the number of factors required further analysis. In (\cite{BroKiers2003})  a test (Core Consistency Diagnostic or CORCONDIA) is proposed to determine whether, given a fixed number of factors, the model is better explained by a Tucker decomposition (Eq. \ref{TUCKER}) or by a CP
decomposition (Eq. \ref{PARAFAC}). Such an approach tests whether allowing for non--zero values across the off--diagonal elements of the core tensor $\boldsymbol{\mathscr{G}}$ in Equation \ref{TUCKER} gives a better fit for the
model. In our case, we found quite clearly that the CORCONDIA test is consistently very close to one (i.e. a CP model provides a good fit w.r.t. a Tucker model) for a number of factors $R\geq3$. As such, we focused on the
number of factors by taking into account both a good level of fit and parsimony. Figure 2 shows that a good fit is achieved for $R\geq25$. After several attempts, we also noticed that the hubs and authorities scores became
stable for $R\geq30$. As such, combining this information, we decided to set $R=30$ in our analysis, which provides a good fit of $62\%$ while keeping the number of factors relatively low with respect to the dimensionality of the
problem, still allowing for a meaningful sub--groups identification.

Moreover, all hubs/authorities/topic scores tend not to vary for higher valuer of $R$ (especially for the highest--scoring companies).

As previously noted, while BD (i.e. the second layer/slice) is symmetric by construction, SH is asymmetric and, as such, provides the reason for computing separated hubs/authorities scores in the TOPHITS model.

\section{Results}
\subsection{HITS}
A first analysis was carried on the two univariate networks (SH and BD). For the (directed) SH network, hubs/authorities scores have been computed, whereas, for the (undirected) BD network, we have computed the eigencentralities. All results hereby presented (Tables \ref{tablescore1} and \ref{tablescore2}) are computed for each of the SCCs in each layer/network.

\begin{table}[ptbh]
\caption{Hubs/Authorities score in the univariate network (first principal vector) for the SH network}%
\label{tablescore1}
\centering \input{univariate_1.tex}
\end{table}
\begin{table}[ptbh]
\caption{Highest eigencentrality nodes in the univariate network (first eigenvector) for the Board network, and their degree}%
\label{tablescore2}%
\centering \input{univariate_2.tex}
\end{table}

It is very clear that the core of shareholding in Italy is composed of banks and financial institutions. Eight out of the ten most eigencentral (hubs) are banks or insurance companies. Fiat and Pirelli are the only exceptions. On the side of held companies, industrial groups emerge, Mediobanca having a prominent role as a holding and held company.

From the BD side, the picture changes completely. The industrial companies play a central role through their directors sitting in different boards. Pirelli, the most degree central and eigencentral, shares its directors with eighteen other companies. Moreover, Mediobanca shows here again its central role.

\subsection{TOPHITS}
As mentioned in \ref{methodology}, the TOPHITS algorithm reveals triplets for hubs, authorities and layer (network) scores. In our case, the network scores refer to either the SH or BD network, i.e. one of the layers of the corporate governance.

As in \cite{KolBadKen2005}, we will first refer to the \emph{dominant} triplet $\{\mathbf{h}^{(1)},\mathbf{a}^{(1)},\mathbf{t}^{(1)}\}$, i.e. the first element ($r=1$) of the sum in Equation \ref{PARAFAC}, as the \emph{dominant grouping}, which refers to the most important corporate governance group and
captures the most relevant control structure in the whole network.

The normalized layer (topic) scores in this case are $0.7582$ for the SH network and $0.2418$ for the Board network. In other terms, the network data structure is explained $75\%$ by the shareholding network and only $24\%$ by the board network. This implies that, notwithstanding the normalization performed in the data preprocessing, a much sparser structure (SH) carries much more topological information than a denser one (BD).

A simple, yet interesting, economic interpretation of this is that the multiple network of corporate governance has a very characteristic topology in its SH component which, although sparse, has a fundamental importance in
connecting the most relevant companies, and the much denser Board network provides the actual linkages backbone amongst companies. Such a backbone structure is denser yet less significant in capturing the most relevant ties
in corporate governance.

The hubs/authorities for the dominant grouping in the TOPHITS model reveal interesting information as well. By taking into account the multiple ties amongst companies, those companies that have a systemic governance influence
(also via indirect ties along different dimensions) are more likely to emerge.

\begin{table}[ptb]
\caption{Hubs/Authorities score (TOPHITS) (2 layers, $r = 1$)}
\label{tablescore3}
\centering \input{table_score.tex}
\end{table}

Table \ref{tablescore3} show the $10$ highest ranking companies in both hubs and authorities scores for $r=1$. The rankings are quite interesting, as companies that are well known for being extremely relevant in the Italian corporate governance structure rank high. In particular, Mediobanca ranks high in both hub and authority scores, because it
is both a central player in controlling other companies. The chain of shareholding control of Fondiaria--Sai is of particular interest. In the univariate SH network, Fondiaria ranks first because it owns the $3.14\%$ of
Mediobanca, $2.24\%$ of RCS Mediagroup and $4.20\%$ of Gemina. However, in the bivariate network Fondiaria--Sai completely disappears from the highest scoring companies, while Mediobanca reemerges as the key hub of the network, thanks to its predominance in \emph{both} network layers. On the other hand, for other companies, the effect of boards on ownership is somehow limited. The central companies in the univariate SH network remain central in the TOPHITS
model. Exceptions are Mediobanca, Intesa San Paolo and UniCredit, whose large boards structure make them more relevant. Furthermore, taking into account both networks simultaneously, the influence of non--listed companies (not present in the SH network) but with representatives in the board (present on the BD network) can be revealed. This is the case of Caltagirone and Premafin, both family companies. Further assessment of the impact of holding weights will be the aim of further research.

Since the SH network is the one ranking highest in terms of topic score (i.e. it carries most topological structure) and it is asymmetric, it is the only layer/slice providing different hub/authority scores. Should, in fact, the two layers be both symmetric, the hub and authority scores would coincide.

In other words, hubs and authorities are correctly found accruing to the ownership. From Table 3, we see once more how the Italian ownership structure is dominated by banks and, in particular, by Mediobanca. RCS Mediagroup is a clear example of the many links among private companies and banks. Among its main shareholders we find: Banco Popolare di Sondrio, Fiat Spa, Fondiaria-Sai, Intesa San Paolo, Banca Italmobiliare, Mediolanum Spa, Milano Assicurazioni, Pirelli \& Co.

\subsection{Sub--groups}
As previously explained, the TOPHITS algorithm allows for the identification of sub--groups of nodes in the multiple network structure. In this light, we checked the scores the different values of $r = 1,2,3, \ldots, 30$, identifying
interesting sub--groups of companies that are notoriusly known to lie within the same sphere of control on both shareholding and board structure. In particular, we found (see Tables \ref{sub1}, \ref{sub2}, \ref{sub3} and
\ref{sub4}) interesting sub--groups for $r=6$ (Caltagirone), $r=9$ (FIAT group), $r=12$ (Poligrafici) and $r=17$ (ENI).


Some important industrial sectors are represented: Caltagirone, active in the construction sector and owns Vianini and Cementir is connected to the whole system through Unicredit and Mediobanca. Poligrafici Ed. and the controller Monrif operate in publishing, printing and new media. Eni, an ex government owned company is stil with a strong public participation, as well as Finmeccanica. They operate in the sector of energy and mechanics, as well as most of the companies they control.

\begin{table}[tbh]
\caption{TOPHITS model (2 layers, $r = 6. $) [Caltagirone]}
\label{sub1}
\centering \input{table_score_6_factor.tex}
\end{table}
\begin{table}[tbh]
\caption{TOPHITS model (2 layers, $r = 9. $) [FIAT] }%
\label{sub2}
\centering \input{table_score_9_factor.tex}
\end{table}
\begin{table}[tbh]
\caption{TOPHITS model (2 layers, $r = 12. $) [Poligrafici] }%
\label{sub3}
\centering \input{table_score_12_factor.tex}
\end{table}
\begin{table}[tbh]
\caption{TOPHITS model (2 layers, $r = 17. $) [ENI]}%
\label{sub4}%
\centering \input{table_score_17_factor.tex}
\end{table}

\subsection{Governance and prices}
As showed above, the corporate governance topological structure can be quantified via assigning companies scores that capture their governance relevance in the multiple network. We now consider also the third layer: the return on prices correlation network, as described above. The new multiple network is now composed of three layers and we again computed, analogously to what we have previously done, the TOPHITS score.

As a first result, we noticed that the same number of factors $R=30$ leads to having a similar fit $61\%$ with respect to the two--layer case, but the topic/layer scores change dramatically: the SH network has a normalized score
of $0.15$, whereas the board network has a very low score of $0.04$; the score of the correlation network is very high $0.81$. This shows that significant price correlations have a very structured topology, which can be due to the
contingent situation of the financial crisis (prices tend to co--move more in the same direction). The striking result, however, is that the corporate governance structure can explain very little of a company performance on the
stock market.

New hubs/authorities scores can be computed (Table \ref{tablescore4}) and they show a much higher degree of symmetry by stressing even more the important role of banks in the Italian network.

New companies enter among the most relevant, ENEL Spa and Exor, a financial holding controlled by the Agnelli family. Apparently, Enel and Exor are among the leaders not only in terms of ownership, but also as drivers of market prices.

\begin{table}[!htbp]
\caption{Hubs/Authorities score in the TOPHITS model ($3$ layers)}
\label{tablescore4}%
\centering \input{table_score2.tex}
\end{table}

\renewcommand{\theenumi}{\roman{enumi}}

\section{Conclusions and further research}
The univariate and multivariate analyses, carried on in this paper, have brought to the following results:
\begin{enumerate}
\item As far as ownership is concerned, the core of shareholding is composed of banks and financial institutions. Eight out of the ten most relevant companies are banks or insurance companies. Fiat and Pirelli \& Co. are the only exceptions.
\item From the board of directors network side, the most central ones are industrial companies: Pirelli, the most central, shares its directors with eighteen other companies.

\item A few industrial groups emerge: Pirelli, Fiat, Caltagirone, Premafin, Poligrafici, Enel, Finmeccanica and ENI. These groups are family companies or ex-state owned companies with a still strong public participation.
\item Mediobanca is definitely the Italian leader in CG, present in all industrial and financial groups, or at least the most relevant in terms of participation.
\item The effect of CG on market performance is practically null. A common driver, probably due to the strong economic crisis as of April 2012 -- April 2013, is leading the market.
\item A further time analysis could possibly show whether this scarce relevance of shareholding and boards on the market performance is contingent phenomenon or, on the contrary, is persistent.
\end{enumerate}

The results show clearly that the successful way to analyze complex interrelations is by multi--way analysis. 
In particular, corporate governance cannot be understood unless ownership and management are taken simultaneously. 
In multi--way analysis, more modes can be added, like we did with price correlations, thus shading new light on 
the relationship between management and control and the market. More research can be done introducing the time effect, 
thereby having a clearer picture of the evolution of power.

\end{document}

%% file: abstract.tex
In this work, we consider Corporate Governance (CG) ties among companies from a multiple network perspective. Such a structure naturally arises from the close interrelation between the Shareholding Network (SH) and the Board of Directors network (BD).  In order to capture the simultaneous effects of both networks on CG,  we propose to model the CG multiple network structure via tensor analysis. In particular, we consider the TOPHITS model, based on the PARAFAC tensor decomposition, to show that tensor techniques can be successfully applied in this context. By providing some empirical results from the Italian financial market in the univariate case, we  then show that a tensor--based multiple network approach can reveal important information.

%% file: univariate_1.tex
\begin{small}\begin{tabular}{|c|c|c|c|}
\hline
\textbf{company}&\textbf{hubs score}&\textbf{company}&\textbf{authority score}\\\hline
Fondiaria-Sai&0.17417&Rcs Mediagroup&0.16885\\\hline
Unicredit Spa&0.12595&Mediobanca&0.12645\\\hline
Italmobiliare&0.11921&Risanamento Spa&0.11343\\\hline
Banco Popolare S&0.1026&Gemina Spa&0.10322\\\hline
Intesa Sanpaolo&0.1026&Industria E Inno&0.080684\\\hline
Banca Monte Dei&0.08252&Milano Assicuraz&0.047759\\\hline
Mediobanca&0.076291&Fondiaria-Sai&0.03798\\\hline
Fiat Spa&0.061368&Alerion&0.032925\\\hline
Pirelli \& C.&0.055806&Italcementi&0.032688\\\hline
Mediolanum Spa&0.041792&Mittel Spa&0.032688\\\hline
\end{tabular}
\end{small}

%% file: univariate_2.tex
\begin{small}\begin{tabular}{|c|c|c|}
\hline
\textbf{company}&\textbf{eigencentrality}&\textbf{degree}\\\hline
Pirelli \& C.&0.034224&18\\\hline
Mediobanca&0.033053&17\\\hline
Italcementi&0.032149&18\\\hline
Atlantia Spa&0.02823&16\\\hline
Luxottica Group&0.023417&15\\\hline
Generali Assic&0.020997&15\\\hline
Fiat Industrial&0.020907&11\\\hline
Brembo Spa&0.019704&11\\\hline
Autogrill Spa&0.018285&9\\\hline
Italmobiliare&0.018&9\\\hline
\end{tabular}
\end{small}

%% file: table_score.tex
\begin{small}\begin{tabular}{|c|c|c|c|}
\hline
\textbf{company}&\textbf{hubs score}&\textbf{company}&\textbf{authority score}\\\hline
Mediobanca&0.17674&Rcs Mediagroup&0.16773\\\hline
Banco Popolare S&0.16796&Mediobanca&0.10579\\\hline
Intesa Sanpaolo&0.15786&Gemina Spa&0.096294\\\hline
Italmobiliare&0.15212&Risanamento Spa&0.068168\\\hline
Unicredit Spa&0.14603&Mittel Spa&0.049657\\\hline
Fiat Spa&0.095423&Italcementi&0.046879\\\hline
Pirelli \& C.&0.07785&Piquadro Spa&0.043704\\\hline
Premafin Finanz&0.040173&Generali Assic&0.039667\\\hline
Cattolica Assic&0.029926&Pirelli \& C.&0.039256\\\hline
Caltagirone Spa&0.023077&Dea Capital Spa&0.037108\\\hline
\end{tabular}
\end{small}

%% file: table_score_6_factor.tex
\begin{small}\begin{tabular}{|c|c|c|c|}
\hline
\textbf{company}&\textbf{hubs score}&\textbf{company}&\textbf{authority score}\\\hline
Caltagirone Spa&0.16302&Cementir Holding&0.15222\\\hline
Caltagirone Edit&0.14021&Vianini Industri&0.15214\\\hline
Unicredit Spa&0.07993&Vianini Lavori&0.15214\\\hline
Fondiaria-Sai&0.058213&Caltagirone Edit&0.079104\\\hline
Mediobanca&0.055692&Italmobiliare&0.027755\\\hline
\end{tabular}
\end{small}

%% file: table_score_9_factor.tex
\begin{small}\begin{tabular}{|c|c|c|c|}
\hline
\textbf{company}&\textbf{hubs score}&\textbf{company}&\textbf{authority score}\\\hline
Fiat Spa&0.052963&Fiat Industrial&0.048148\\\hline
Mediobanca&0.042861&Fiat Spa&0.034534\\\hline
Rcs Mediagroup&0.036761&Fondiaria-Sai&0.03451\\\hline
Fondiaria-Sai&0.033111&Brembo Spa&0.028077\\\hline
Intesa Sanpaolo&0.027334&Luxottica Group&0.024689\\\hline
\end{tabular}
\end{small}

%% file: table_score_12_factor.tex
\begin{small}\begin{tabular}{|c|c|c|c|}
\hline
\textbf{company}&\textbf{hubs score}&\textbf{company}&\textbf{authority score}\\\hline
Poligrafici Prin&0.27396&Poligrafici Edit&0.31366\\\hline
Poligrafica San&0.27394&Poligrafici Prin&0.24499\\\hline
Poligrafici Edit&0.26587&Poligrafica San&0.24497\\\hline
Monrif Spa&0.14185&Noemalife Spa&0.010925\\\hline
Finmeccanica Spa&0.0045832&Bolzoni Spa&0.010913\\\hline
\end{tabular}
\end{small}

%% file: table_score_17_factor.tex
\begin{small}\begin{tabular}{|c|c|c|c|}
\hline
\textbf{company}&\textbf{hubs score}&\textbf{company}&\textbf{authority score}\\\hline
Eni Spa&0.17295&Saipem Spa&0.14147\\\hline
Italmobiliare&0.075896&Snam Spa&0.13879\\\hline
Intesa Sanpaolo&0.071711&Ansaldo Sts Spa&0.059603\\\hline
Finmeccanica Spa&0.066764&Eurotech Spa&0.059328\\\hline
Premafin Finanz&0.055171&Rcs Mediagroup&0.050548\\\hline
\end{tabular}
\end{small}

%% file: table_score2.tex
\begin{small}\begin{tabular}{|c|c|c|c|}
\hline
\textbf{company}&\textbf{hubs score}&\textbf{company}&\textbf{authority score}\\\hline
Intesa Sanpaolo&0.044282&Intesa Sanpaolo&0.043007\\\hline
Unicredit Spa&0.043979&Generali Assic&0.042384\\\hline
Generali Assic&0.042652&Unicredit Spa&0.042056\\\hline
Mediolanum Spa&0.042345&Mediolanum Spa&0.041021\\\hline
Banca Pop Emilia&0.041979&Banca Pop Emilia&0.04018\\\hline
Ubi Banca Scpa&0.041854&Ubi Banca Scpa&0.039772\\\hline
Enel Spa&0.040269&Enel Spa&0.038583\\\hline
Mediobanca&0.038018&Exor&0.037093\\\hline
Exor&0.038014&Eni Spa&0.035045\\\hline
Banca Pop Sondri&0.035452&Atlantia Spa&0.034714\\\hline
\end{tabular}
\end{small}